\documentclass[10pt,journal]{IEEEtran}

\ifCLASSINFOpdf
	\usepackage{graphicx}
\else
	\usepackage[dvips]{graphicx}
\fi
 \usepackage{ifthen}
 \newboolean{bookver}
 \setboolean{bookver}{false}

\usepackage[latin1]{inputenc} 
\usepackage[sort&compress]{natbib}
\usepackage[cmex10]{amsmath} 
\usepackage{array} 
\usepackage{fixltx2e} 
\usepackage{amssymb} 
\usepackage{amsthm} 
\usepackage{dsfont} 
\usepackage{bm} 
\usepackage[printonlyused]{acronym} 
\usepackage{color} 
\usepackage{hyperref}
\usepackage[all]{hypcap}
\usepackage{cleveref}
\usepackage{epstopdf}
\usepackage{multirow}
\usepackage{booktabs}
\usepackage{bigstrut}

\usepackage{enumitem}

 \bibpunct{[}{]}{,}{n}{,}{;}

\definecolor{myBlue}{rgb}{0,0,0.55}
 \definecolor{xBlue}{rgb}{0,0,0.5}
 \hypersetup{
  linkcolor=  myBlue,
  citecolor=  myBlue,
  urlcolor=   myBlue,
  pdffitwindow=       true,
  breaklinks=         true,
  pdfcenterwindow=    true,
  pdfstartview=       FitH,
  bookmarks=          false,
  bookmarksopen=      false,   
  bookmarksnumbered=  false,   
  pdfhighlight=       /P,
  linktocpage=        true,   
  colorlinks=         false,
  pdfborder=          0 0 0,
  pdftitle=           {Pushing the Limits of LTE: A Survey on Research Enhancing the Standard},
  pdfauthor=          {Stefan Schwarz, Josep Colom Ikuno, Michal \v{S}imko, Martin Taranetz, Qi Wang, and Markus Rupp},
  pdfkeywords=        {LTE, MIMO, reproducible research, heterogeneous networks, distributed antenna systems, femto cells, frequency synchronization, pilot power allocation, multi-user gains},
  }

\usepackage[switch]{lineno}
\newcommand*\patchAmsMathEnvironmentForLineno[1]{%
 \expandafter\let\csname old#1\expandafter\endcsname\csname #1\endcsname
 \expandafter\let\csname oldend#1\expandafter\endcsname\csname end#1\endcsname
 \renewenvironment{#1}%
    {\linenomath\csname old#1\endcsname}%
    {\csname oldend#1\endcsname\endlinenomath}}%
\newcommand*\patchBothAmsMathEnvironmentsForLineno[1]{%
 \patchAmsMathEnvironmentForLineno{#1}%
 \patchAmsMathEnvironmentForLineno{#1*}}%
\AtBeginDocument{%
\patchBothAmsMathEnvironmentsForLineno{equation}%
\patchBothAmsMathEnvironmentsForLineno{align}%
\patchBothAmsMathEnvironmentsForLineno{flalign}%
\patchBothAmsMathEnvironmentsForLineno{alignat}%
\patchBothAmsMathEnvironmentsForLineno{gather}%
\patchBothAmsMathEnvironmentsForLineno{multline}%
}

\newcommand{\mat}{\textsc{Matlab}} 

\acrodef{3GPP}[3GPP]{3rd Generation Partnership Project}
\acrodef{ACK}[ACK]{ACKnowledged}
\acrodef{AMC}[AMC]{Adaptive Modulation and Coding}
\acrodef{ARQ}[ARQ]{Automatic Repeat reQuest}
\acrodef{AWGN}[AWGN]{Additive White Gaussian Noise}
\acrodef{BER}[BER]{Bit Error Ratio}
\acrodef{BICM}[BICM]{bit-interleaved coded modulation}
\acrodef{BLER}[BLER]{BLock Error Ratio}
\acrodef{CDF}[CDF]{Cumulative Density Function}
\acrodef{CDMA}[CDMA]{Code-Division Multiple Access}
\acrodef{CFO}[CFO]{carrier frequency offset}
\acrodef{CoMP}[CoMP]{Cooperative Multi-Point}
\acrodef{CPICH}[CPICH]{Common Pilot CHannel}
\acrodef{CQI}[CQI]{Channel Quality Indicator}
\acrodef{CRC}[CRC]{Cyclic Redundancy Check}
\acrodef{CSI}[CSI]{Channel State Information}
\acrodef{CSIT}[CSIT]{channel state information at the transmitter}
\acrodef{DAS}[DAS]{distributed antenna system}
\acrodef{DoF}[DoF]{Degrees of Freedom}
\acrodef{DTxAA}[D-TxAA]{Double Transmit Antenna Array}
\acrodef{ECR}[ECR]{Effective Code Rate}
\acrodef{EESM}[EESM]{Exponential Effective \acl{SINR} Mapping}
\acrodef{FDD}[FDD]{Frequency Division Duplex}
\acrodef{FFT}[FFT]{Fast Fourier Transform}
\acrodef{GSM}[GSM]{Global System for Mobile communications}
\acrodef{HARQ}[HARQ]{Hybrid \acl{ARQ}}
\acrodef{HSDPA}[HSDPA]{High-Speed Downlink Packet Access}
\acrodef{HSPA}[HSPA]{High-Speed Packet Access}
\acrodef{HSDPCCH}[HS-DPCCH]{High-Speed Dedicated Physical Control CHannel}
\acrodef{HSDSCH}[HS-DSCH]{High-Speed Downlink Shared CHannel}
\acrodef{HSPDSCH}[HS-PDSCH]{High-Speed Physical Downlink Shared CHannel}
\acrodef{HSSCCH}[HS-SCCH]{High-Speed Shared Control CHannel}
\acrodef{HSUPA}[HSUPA]{High-Speed Uplink Packet Access}
\acrodef{IMS}[IMS]{IP Multimedia Subsystem}
\acrodef{ICI}[ICI]{Inter Carrier Interference}
\acrodef{ISI}[ISI]{Inter Symbol Interference}
\acrodef{ITU}[ITU]{International Telecommunication Union}
\acrodef{LEP}[LEP]{Link Error Prediction}
\acrodef{LS}[LS]{Least Squares}
\acrodef{LTE}[LTE]{Long Term Evolution}
\acrodef{LTEA}[LTE-A]{\acs{LTE} Advanced}
\acrodef{LMMSE}[LMMSE]{Linear Minimum Mean Squared Error}
\acrodef{MAChs}[MAC-hs]{Medium Access Control for \acl{HSDPA}}
\acrodef{MCS}[MCS]{Modulation and Coding Scheme}
\acrodef{MIESM}[MIESM]{Mutual Information Effective \acl{SINR} Mapping}
\acrodef{MIMO}[MIMO]{multiple-input multiple-output}
\acrodef{ML}[ML]{Maximum Likelihood}
\acrodef{MMSE}[MMSE]{minimum mean squared error}
\acrodef{MSE}[MSE]{mean squared error}
\acrodef{MU}[MU]{Multi User}
\acrodef{MUMIMO}[MU-MIMO]{multi-user \acs{MIMO}}
\acrodef{MVU}[MVU]{Minimum Variance Unbiased}
\acrodef{NACK}[NACK]{Non-ACKnowledged}
\acrodef{PCI}[PCI]{Precoding Control Information}
\acrodef{PMI}[PMI]{Precoding Matrix Indicator}
\acrodef{PedA}[PedA]{Pedestrian A}
\acrodef{PedB}[PedB]{Pedestrian B}
\acrodef{PU2RC}[PU2RC]{per-user unitary beamforming and rate control} 
\acrodef{PHY}[PHY]{Physical}
\acrodef{QoS}[QoS]{Quality of Service}
\acrodef{RAN}[RAN]{Radio Access Network}
\acrodef{RB}[RB]{Resource Block}
\acrodef{RI}[RI]{Rank Indicator}
\acrodef{ROI}[ROI]{Region of Interest}
\acrodef{RLC}[RLC]{Radio Link Control}
\acrodef{RRC}[RRC]{Radio Resource Control}
\acrodef{RRU}[RRU]{remote radio unit}
\acrodef{UE}[UE]{user equipment}
\acrodef{OFDMA}[OFDMA]{Orthogonal Frequency Division Multiple Access}
\acrodef{OFDM}[OFDM]{Orthogonal Frequency Division Multiplexing}
\acrodef{PDP}[PDP]{Power Delay Profile}
\acrodef{SAE}[SAE]{System Architecture Evolution}
\acrodef{SCM}[SCM]{Spatial Channel Model}
\acrodef{SISO}[SISO]{single-input single-output}
\acrodef{SM}[SM]{Spatial Multiplexing}
\acrodef{SINR}[SINR]{signal to interference and noise ratio}
\acrodef{SNR}[SNR]{signal to noise ratio}
\acrodef{SQP}[SQP]{Sequential Quadratic Programming}
\acrodef{STMMSE}[ST-MMSE]{Space-Time \acl{MMSE}}
\acrodef{SU}[SU]{Single User}
\acrodef{SUMIMO}[SU-MIMO]{single-user \acs{MIMO}}
\acrodef{SVD}[SVD]{singular value decomposition}
\acrodef{TB}[TB]{Transport Block}
\acrodef{TBS}[TBS]{Transport Block Size}
\acrodef{TTI}[TTI]{Transmission Time Interval}
\acrodef{TxAA}[TxAA]{Transmit Antenna Array}
\acrodef{UMTS}[UMTS]{Universal Mobile Telecommunications System}
\acrodef{UTRA}[UTRA]{\acl{UMTS} Terrestrial Radio Access}
\acrodef{WCDMA}[W-CDMA]{Wideband \acl{CDMA}}
\acrodef{ZF}[ZF]{zero forcing}
\acrodef{CP}[CP]{Cyclic Prefix}
\acrodef{RB}[RB]{Resource Block}
\acrodef{PDSCH}[PDSCH]{Physical Downlink Shared Channel}
\acrodef{QPSK}[QPSK]{Quadrature Phase Shift Keying}
\acrodef{VehA}[VehA]{Vehicular A}
\acrodef{EVehA}[EVehA]{Extended \acl{VehA}}
\acrodef{CLSM}[CLSM]{closed-loop spatial multiplexing}
\acrodef{OLSM}[OLSM]{Open Loop Spatial Multiplexing}
\acrodef{TxD}[TxD]{Transmit Diversity}
\acrodef{SSD}[SSD]{Soft Sphere Decoder}
\acrodef{WiMAX}[WiMAX]{Worldwide Interoperability for Microwave Access}

\crefformat{chapter}{#2Chapter~#1#3}
\crefformat{lem}{#2Lemma~#1#3}
\crefformat{section}{#2Section~#1#3}
\crefformat{subsection}{#2Section~#1#3}
\crefformat{appendix}{#2Appendix~#1#3}
\crefformat{figure}{#2Figure~#1#3}
\crefformat{table}{#2Table~#1#3}
\crefformat{equation}{#2Equation~(#1)#3}
\crefmultiformat{chapter}{#2Chapters~#1#3}{ and~#2#1#3}{, ~#2#1#3}{, and~#2#1#3}
\crefmultiformat{section}{#2Sections~#1#3}{ and~#2#1#3}{, ~#2#1#3}{, and~#2#1#3}
\crefmultiformat{equation}{#2Equations~(#1)#3}{ and~#2(#1)#3}{, ~#2(#1)#3}{, and~#2(#1)#3}
\crefmultiformat{figure}{#2Figures~#1#3}{ and~#2#1#3}{, ~#2#1#3}{, and~#2#1#3}
\crefmultiformat{table}{#2Tables~#1#3}{ and~#2#1#3}{, ~#2#1#3}{, and~#2#1#3}
\crefrangeformat{equation}{#3Equations~(#1)#4--#5(#2)#6}
\crefrangeformat{figure}{#3Figures~#1#4--#5#2#6}
\crefrangeformat{chapter}{#3Chapters~#1#4 to #5#2#6}

\begin{document}
\title{Pushing the Limits of LTE: A Survey on Research Enhancing the Standard}

\author{Stefan Schwarz, Josep Colom Ikuno, Michal \v{S}imko, Martin Taranetz, Qi Wang, and Markus Rupp%
\thanks{S. Schwarz, J. Colom Ikuno, M. \v{S}imko, M. Taranetz, Q. Wang, and M. Rupp are with the Institute of Telecommunications, Vienna University of Technology, Austria; email: \{sschwarz, jcolom, msimko, mtaranet, qwang, mrupp\}@nt.tuwien.ac.at}%
}


\maketitle

\begin{minipage}{500pt}
\vspace{-350pt}
\centering{
\footnotesize{Copyright 2013 IEEE, Accepted for publication in IEEE Access}}
\end{minipage}

\begin{abstract} 
Cellular networks are an essential part of todays communication infrastructure. The ever-increasing demand for higher data-rates calls for a close cooperation between researchers and industry/standardization experts which hardly exists in practice. In this article we give an overview about our efforts in trying to bridge this gap. Our research group provides a standard-compliant open-source simulation platform for 3GPP LTE that enables reproducible research in a well-defined environment. We demonstrate that much innovative research under the confined framework of a real-world standard is still possible, sometimes even encouraged. With examplary samples of our research work we investigate on the potential of several important research areas under typical practical conditions.
\end{abstract}

\acresetall 

\section{Introduction}


Life without ubiquitous possibilities to connect to the Internet is hard to imagine nowadays. 
Cellular networks play a central role in this global networking and communication infrastructure. To ensure and even enhance availability, the standardization process of new communication systems is governed by concerns about reliability, interoperability and security, besides trying to improve the performance of current technology. Still, the ever-increasing demand for higher data-rates forces the consideration of novel research results during standardization. For good reason, however, standardization experts are reluctant about innovative results: Often assumptions made by researchers are too simplistic and idealistic to reflect the performance under practical conditions.

The world's leading cellular networking technology these days is standardized by the \ac{3GPP}, a collaboration between telecommunication associations spread all over the world. Technical specifications for the radio access network technology, the core network and the service architecture are released every few years, constantly evolving the cellular system with a major focus on compatibility between releases. With the introduction of \ac{LTE} in Release~8 (2008), an entirely new air interface based on \ac{OFDM} was implemented setting the basis for a 4G capable mobile communication technology, and first \ac{LTE} networks went on-air in 2009/10. Since then, work on \ac{LTEA} and beyond is ongoing in the 3GPP and the corresponding research communities.

In trying to classify research work with respect to its practical applicability, theoretical results mostly lag behind due to the coarse abstraction required to facilitate analytical tractability. Such results have their significance in providing upper bounds on system performance and opening new fields for research and engineering activities. Simulations enable investigations of more complex and detailed scenarios, and facilitate comparison of different algorithms under identical conditions. Measurements and field trials avoid all kinds of assumptions and models, thus reflecting reality most closely. Still, the involved expenditure of time, labor and money to carry out measurement campaigns, and the lack of generality in the obtained results prohibits their application in early stages of research.   

The simulation approach is adopted by many researchers in combination with theoretical investigations, due to its flexibility and efficiency. Though standardized simulation models exist for parts of the environment, e.g., the wireless channel in cellular communications, there is still lots of ambiguity in many simulation parameters left, making it often difficult to reproduce results of other researchers and hampering cross-comparison of different techniques. Moreover, the code of such highly complex systems may contain more than 100.000 lines, making thorough testing practically impossible in a reasonably short time. Only working in parallel with many independent research groups and communicating publicly via a web-based forum makes it possible to identify programming bugs and have the code checked independently several times.
These facts were the motivation for our research group to develop standard-compliant open-source \mat-based link- and system-level simulation environments for \ac{LTE}, namely the \textit{Vienna LTE simulators}~\cite{ViennaLTESimulators}. The simulators are publicly available for download (\href{http://www.nt.tuwien.ac.at/ltesimulator/}{www.nt.tuwien.ac.at/ltesimulator}) under an academic non-commercial use license. They facilitate reproducibility of research results, and contribute to bridge the gap between researchers and standardization experts in \ac{LTE} and \ac{LTEA}. Since its first release in 2009, the \ac{LTE} downlink link-level simulator was downloaded more than 13\,000 times and is currently (August 2012) in its eighth release. It was also extended to \ac{LTEA} and augmented with an uplink version in 2011. The \ac{LTE} system-level simulator experienced similar attention, with more than 17\,000 downloads, thus confirming the demand for a consistent simulation environment.

\begin{figure}[t]
\centering
\includegraphics[scale = 0.85]{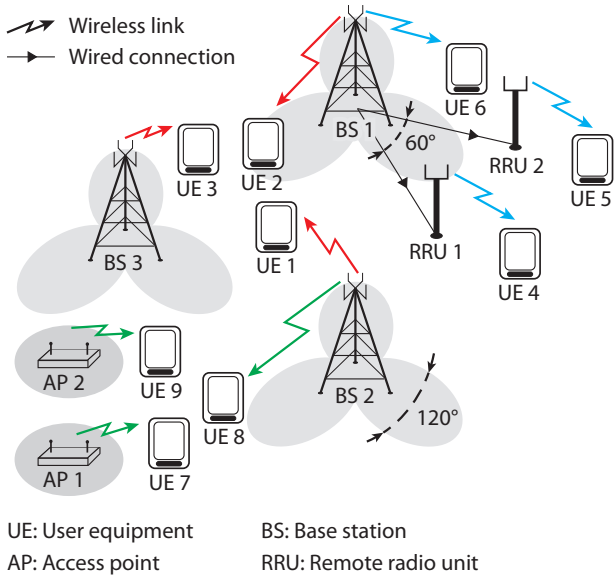}
\caption{Example cellular network consisting of three sectorized macro base stations and additional radio access equipment, visualizing different scenarios considered in our research work.}
\label{fig:Network}
\end{figure}

In our research work, the simulators are utilized to investigate cellular networks in varying degrees of abstraction. Consider the example cellular network shown in~\cref{fig:Network}. It consists of three macro base stations with sectorized antennas, plus additional radio access equipment. Users 1--3 are served in the ''classical'' way, by attaching the \ac{UE} to the strongest macro base station and treating other base stations as interferers. The data-transmission can be optimized by focusing on the radio link between a base station and a single \ac{UE}, which relies on detailed modeling of the physical-layer and requires link-level simulations. Two examples for link-optimization are treated in~\cref{sec:Qi,sec:Michal}, on the subjects of pilot power allocation for channel estimation and carrier frequency synchronization.  

An alternative perspective for optimization of cellular communication systems is the network viewpoint. Here, a large network consisting of a multitude of base stations and \acp{UE} is considered. To keep the computational complexity of the associated system-level simulations tractable, abstraction of the physical-layer details is necessary. \cref{sec:Josep} treats multi-user scheduling as an example, confirming the theoretically well-known double-logarithmic growth of the sum-rate with the number of users under the practical constraints that are introduced by the \ac{LTE} standards.    

Extensions of the classical sectorized cellular network architecture to heterogeneous networks, containing different types of radio access equipment, are in the scope of many recent research activities. Two examples are shown in~\cref{fig:Network}. Users 4--6 are jointly served by a single base station whose transmission capabilities are enhanced by \acp{RRU}. The performance of different transmission strategies in such distributed antenna systems is evaluated in~\cref{sec:Stefan}. Femto cell access points are a popular technique for increasing the spatial reuse of existing cellular networks. Users close to access points are offloaded from the macro cellular network and served by the femto cells (see Users 7--9 in \cref{fig:Network}). The benefits of \ac{LTE} femto cell-enhanced macro networks in terms of user throughput and fairness are investigated in~\cref{sec:Martin}, by means of system-level simulations.  

The link-level simulator can also be beneficially employed to simplify measurement campaigns. In our research group it serves as a front-end for a measurement testbed, generating the base band transmit signal and detecting the received signal.


\section{Enhancing the Physical-Layer}
\label{sec:PHY}

Recent measurement and simulation based investigations of current cellular communication systems (HSDPA, WiMAX, LTE) have revealed a large performance gap between the throughput achieved in such systems and the theoretical upper bounds determined by channel capacity~\cite{Caban2012}. 
Although it is often believed that the potential of the physical-layer is already largely exploited, these investigations show that there is still lots of space for improvement left.

Conducting research on the physical-layer of a wireless communication system requires detailed modeling of the wireless channel and the signal-processing algorithms applied at the transmitter and receiver. In this section, two examples of our research work serve to establish the utility of the Vienna LTE link-level simulator for such investigations. In~\cref{sec:Michal} substantial power savings are demonstrated at high user velocities by exploiting the \ac{MSE} saturation of well-known channel estimation algorithms in weakly correlated channels. Furthermore, the impact of a frequency estimation error, causing a carrier frequency offset between transmitter and receiver, on the throughput of an \ac{LTE} system is considered in \cref{sec:Qi}.  


\subsection{Pilot Power Allocation}
\label{sec:Michal}

Modern standards for wireless communication systems such as \ac{LTE} and \ac{WiMAX}  exclusively rely on coherent transmission techniques.
Detection of coherently transmitted data symbols requires knowledge of the channel experienced during transmission, which is obtained by channel estimation. The estimates are calculated from known symbols, so called pilot symbols, that are multiplexed within the data symbols. The amount of power assigned to the pilot symbols has a crucial impact on the quality of the channel estimation, which in turn significantly impacts the throughput performance of the system. The channel estimation error leads to an additional noise term in the \ac{SINR} of the transmission, whose variance is determined by the applied channel estimator and the pilot symbol density and power.

Consider an \ac{LTE} transmitter which has a certain amount of power available for transmission. The available power is divided between pilot and data symbols. The quality of the channel estimates improves with the amount of power invested into the pilot symbols, thus reducing the noise contribution of the channel estimator. But in turn the power of the data symbols has to be decreased to keep the power budget balanced, degrading the received signal power. Therefore, an equilibrium point in the power assignment between pilot and data symbols exists, in which the \ac{SINR} is maximized.


\begin{figure}[t]
\centering
\includegraphics[scale = 0.8]{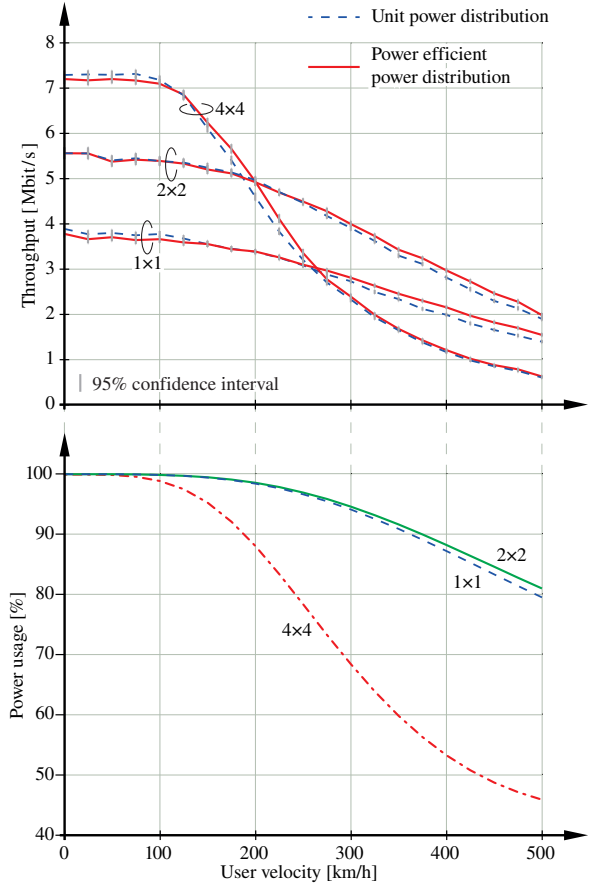}
\caption{Comparison of the average user throughput of an LTE system with unit power allocation for pilot and data symbols, and with power efficient power allocation. The SNR is set to 20\,dB. The lower part indicates the power savings of the power efficient power allocation versus unit power allocation.}
\label{fig:power_alloc}
\end{figure}

The precisely elaborated physical-layer of the Vienna \ac{LTE} link-level simulator enables an extensive investigation of different pilot symbol power allocation algorithms. The optimal distribution of the available transmit power between pilot and data symbols was derived in~\cite{VTC2011_fall_Michal} and~\cite{eurasip_michal_2012} for time-invariant and time-variant channels, respectively. The solution turned out independent of the operating point (\ac{SNR}) and of the actual channel realization, making it very robust and applicable in practical systems. In~\cite{eurasip_michal_2012} it was realized that in time-variant channels state-of-the-art channel estimators (least-squares and linear \ac{MMSE}) exhibit an error-floor, which increases with decreasing temporal channel correlation. Thus, at a given user velocity (which determines the channel correlation in the link-level simulator) and operating point, a further increase of the pilot symbol power does not necessarily lead to an improvement in the quality of the channel estimate. Therefore, one might think that more power should be allocated to the data symbols. This, however, does not improve the post-equalization \ac{SINR} either, because the interlayer interference increases with the data symbol power, due to the imperfect channel knowledge. Based on this insight, a power efficient power allocation algorithm was proposed in~\cite{VTC2012_fall_Michal}. In this algorithm the total transmit power of pilot and data symbols is minimized at a given user velocity and noise power, while constraining the post-equalization \ac{SINR} not to decline with respect to the case that all available transmit power is used (denoted unit power allocation). 

\cref{fig:power_alloc} demonstrates the performance of the power efficient power allocation algorithm in comparison to unit power allocation, as obtained with the link-level simulator. The average user throughput versus user speed for three different antenna configurations $N_t \times N_r \in \left\{1\times 1, 2\times 2,4\times 4 \right\}$ is shown in the upper part of the figure. It can be seen that the power-efficient power allocation algorithm achieves virtually the same throughput as the unit power allocation algorithm. The power usage utilizing the power efficient power allocation algorithm with respect to unit power allocation is depicted in the lower part of \cref{fig:power_alloc}. The figure shows that at high user velocities substantial power savings are possible without degrading the throughput performance of the system. Note that \ac{LTE} is defined to operate up to user velocities of 500\,km/h.

\subsection{Impact of Imperfect Frequency Synchronization}
\label{sec:Qi}

One crucial issue that a novel technique encounters in real world applications is to cope with the physical impairments which are usually not taken into account in simulation-based experiments. For a communication system, typical such examples are an offset between the local oscillators at the transmitter and the receiver, oscillator phase noise or an imbalance between the in-phase and quadrature-phase branches in the front end processing.

\begin{figure}[t]
\centering
\includegraphics[scale = 1.0]{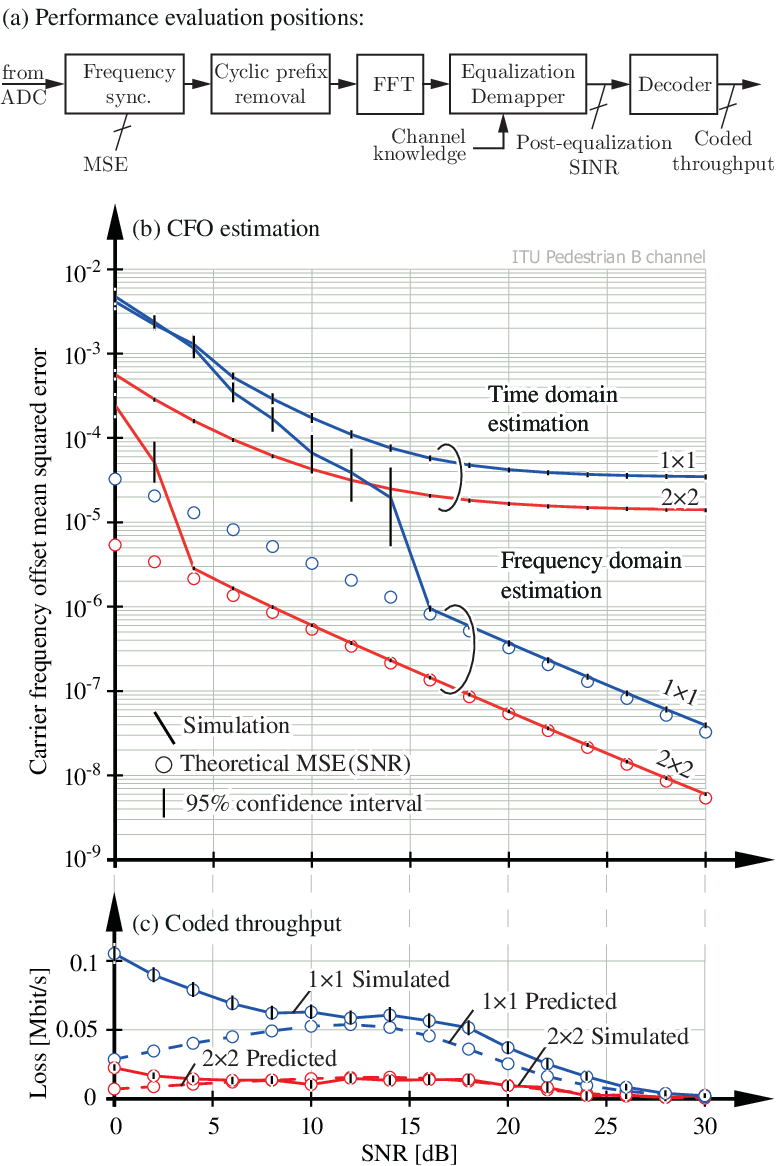}
\caption{(a) The three performance evaluation positions in the receiver signal processing chain. (b) MSE performance of the carrier frequency synchronization scheme in \citep{PIMRC2010_Qi}. (c) The predicted and simulated coded throughput loss resulting from the residual estimation error in (b).}
\label{fig:freqsync}
\end{figure}

Taking the \ac{CFO}, i.e., the offset between the carrier frequencies at the transmitter and the receiver, as an example, plenty of literature can be found on how to estimate the \ac{CFO} in the digital signal processing domain. The estimator's performance is usually evaluated in terms of the estimation error, in other words, the \ac{MSE}. This is shown in the center part of \cref{fig:freqsync} for two specific examples: (i) the time domain and (ii) the frequency domain estimators of \cite{PIMRC2010_Qi}, and two different transmit-receive antenna configurations $N_t \times N_r \in \{1\times 1, 2\times 2\}$. However, the performance of a communication system is evaluated in terms of the overall coded throughput, encompassing all the signal processing steps applied at the transmitter and receiver, e.g., coding, modulation, equalization, and detection. Therefore, a link performance prediction model is desirable, providing a direct mapping from the residual \ac{CFO} to the coded throughput. 
In the following, we show how to utilize such a mapping to estimate the throughput loss caused by the carrier frequency estimation error.

In~\citep{Asilomar2011_Qi}, the authors derive an analytic expression for the post-equalization \ac{SINR} achieved on a resource element\footnote{In \ac{LTE} a resource element denotes the basic unit of physical \ac{OFDM} time-frequency resources.} of the \ac{LTE} downlink with imperfect frequency synchronization. This expression can be exploited to estimate the throughput loss of the \ac{LTE} system, and thus obtain the desired relation:

\begin{enumerate}[leftmargin = 1em]

\item For a given \ac{CFO} estimation scheme, determine the estimation performance in terms of the \ac{MSE}. As shown, e.g., in~\citep{Asilomar2011_Qi} the \ac{MSE} is theoretically given by a function which depends on the \ac{SNR} expressed as $\textnormal{MSE}(\textnormal{SNR})$.

\item Calculate the residual \ac{CFO} $\varepsilon = \sqrt{\textnormal{MSE}(\textnormal{SNR})}$. Utilizing the model of \citep{Asilomar2011_Qi}, the post-equalization \ac{SINR} on a resource element $r$ can then be expressed as a function of the \ac{CFO}, namely $\textnormal{SINR}_r(\varepsilon)$.

\item Estimate the throughput of the system from $\textnormal{SINR}_r(\varepsilon)$. In general, pre-computed mapping tables valid for the considered system (e.g., obtained from link-level simulations) can be employed to map the post-equalization \ac{SINR} to a corresponding throughput value. Here, we are only interested in the throughput loss compared to the case of perfect synchronization. In this case, we can employ the \ac{BICM} capacity to estimate the throughput $f\left(\textnormal{SINR}_r(\varepsilon)\right)$, since \ac{LTE} is based on a \ac{BICM} architecture. The imperfect channel code will cause an offset in the absolute value of the estimated throughput, but this offset approximately cancels out when calculating the throughput loss $\Delta B = \sum_r f\left(\textnormal{SINR}_r(0)\right) - \sum_r f\left(\textnormal{SINR}_r(\epsilon)\right)$. 

\end{enumerate}

The Vienna \ac{LTE} link-level simulator enables such investigations and greatly facilitates a standard compliant validation. In the bottom part of \cref{fig:freqsync}, the predicted loss in terms of the coded throughput is compared to the results obtained by means of extensive link-level simulations. The figure confirms that the prediction model performs well as soon as the \ac{MSE} follows the theoretical $\textnormal{MSE}(\textnormal{SNR})$ relation.

\section{Treating Interference in Cellular Networks}

Many research efforts currently concentrate on the interference between multiple transmitter and receiver pairs. A robust way to deal with interference is opportunistic scheduling, where the interference dynamics are exploited to serve users whenever they experience good channel conditions. This is investigated in~\cref{sec:Josep} by means of system-level simulations. With sufficient \ac{CSIT}, sophisticated signal-processing algorithms can be applied before transmission to avoid/minimize interference between several nodes. In~\cref{sec:Stefan} multi-user beamforming in \acp{DAS} with perfect and quantized \ac{CSIT} is considered, revealing large throughput gains in comparison to \ac{SUMIMO} systems. Finally, in~\cref{sec:Martin} the impact of interference caused by femto cell deployments on existing macro cellular networks is explored with the aid of the Vienna LTE system-level simulator, in terms of achieved user throughputs and resource allocation fairness.    


\subsection{Scheduling and multi-user gain}
\label{sec:Josep}

\begin{figure*}[t]
\centering
\includegraphics[scale = 0.8]{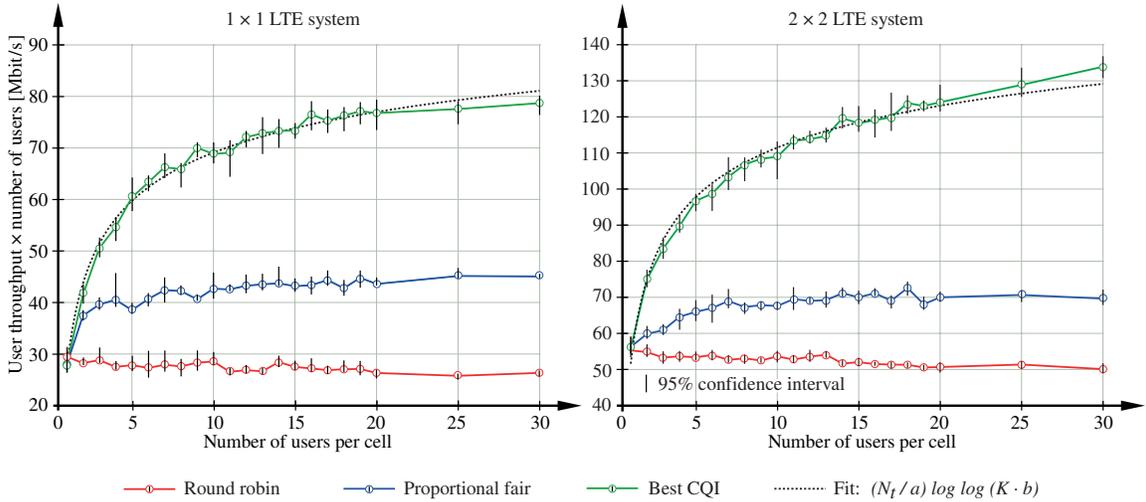}
\caption{Multi-user gain simulation results for a \ac{SISO} (left) and $2\times2$ \ac{CLSM} \acs{LTE} \ac{MIMO} (right) configuration. Solid line: throughput results, Dashed line: $\log \log k$ approximation.}
\label{fig:MU_gain}
\end{figure*}

When a single transmission link is considered, time and frequency diversity can be exploited to increase the throughput and reliability of the data transmission. In a practical cellular network, serving not only a single user but a multitude of them, multi-user diversity can additionally be utilized to increase the total throughput of the cell.
In combination with the spatial degrees of freedom added by a \ac{MIMO} system, theory states that the throughput gain due to multi-user diversity follows an $N_t \log \log k$ rule with the number of concurrent users $k$~\citep{heath2001multiuser}, where $N_t$ is the number of transmit antennas.

The theoretical $N_t \log \log k$ rule, although useful as an upper bound on the achievable diversity, is not directly applicable to the throughput of an \ac{LTE} link, because other parasitic effects encountered in a practical system diminish parts of the promised gains~\citep{Caban2012}, e.g., a growing pilot symbol overhead with increasing number of transmit antennas and a limited choice of possible precoding matrices in \ac{MIMO} systems.   
Hence, there is a need for realistic throughput simulations.

Simulations of large cellular networks, with many cells and users being present, are very computationally demanding. If implemented via link-level simulations, a single simulation of that kind could last several months. By applying physical-layer abstraction models, it is possible to reduce the complexity of such system-level simulations, without significantly compromising the accuracy of the results~\citep{ikuno2010system}. 

\cref{fig:MU_gain} shows simulation results obtained with the Vienna LTE system-level simulator, comparing the performance of several scheduling algorithms in an \ac{LTE} network. The left-hand side results are obtained in a \ac{SISO} system, while the right-hand side performance is achieved in a \ac{MIMO} system with $N_t \times N_r = 2\times2$ employing \ac{LTE}'s \ac{CLSM} mode. The following scheduling schemes are employed:
\begin{enumerate}[leftmargin = 1em]
\item Best CQI scheduling assigns resources to the users with the best channel conditions only. This algorithm is the practical counterpart to the theoretical cell throughput upper bounds, but it achieves the lowest fairness in terms of distributing resources among users.
\item Proportional fair scheduling aims at increasing fairness and avoiding the user starving issue encountered in the best CQI scheduler, by scheduling users whenever their channel conditions are good, compared to their own average channel quality.
\item The round robin scheduler equally distributes resources among users. While the former two algorithms are opportunistic in nature, serving users whenever they experience good channel and interference conditions, this latter algorithm ignores any channel state information and thus does not make use of the available diversity.
\end{enumerate}

An adaptation of the $N_t \log \log k$ rule is employed to quantify the spatial multiplexing and multi-user gains of a practical system, by introducing a scaling factor $a$ for the multiplexing gain and a diversity gain loss factor $b$ for the multi-user diversity.  
The results for the considered \ac{LTE} system show a mutliplexing gain factor ($N_t/a$) of 1.56 for an $N_t \times N_r = 2\times2$ system, and a gain of 2.66 for a $4\times4$ antenna configuration compared to the \ac{SISO} case, thus considerably below the theoretical values of 2 and 4, respectively. A similar analysis applied to proportional fair scheduling shows the same gains relative to the \ac{SISO} case.

Such an analysis of the multi-user gains of \ac{LTE} can serve as the basis for the extension to more complex setups, such as ones including distributed antennas or femto cells.


\subsection{Distributed Antenna Systems}
\label{sec:Stefan}

\begin{figure*}[t]
\centering
\includegraphics[scale = 0.8]{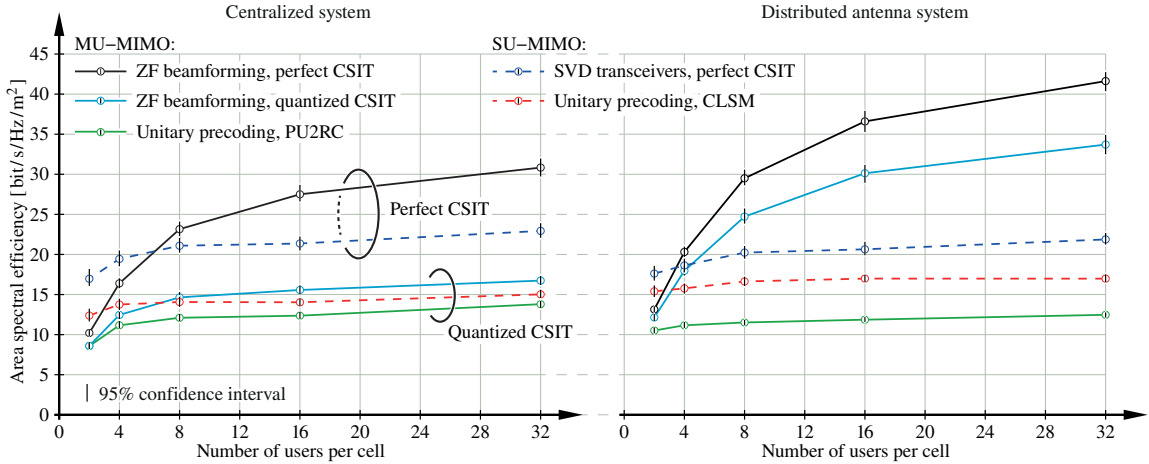}
\caption{Comparison of the area spectral efficiency obtained in a cellular network without (left-hand side) and with (right-hand side) distributed antennas versus the number of users per cell. The performance with different transceivers and \ac{CSI} feedback algorithms is compared. The total amount of transmit antennas per cell equals eight. The users are equipped with four receive antennas.}
\label{fig:ASE}
\end{figure*}

A \ac{DAS} is a cellular networking architecture in which several transmission points, controlled by a single base station, are geographically distributed throughout the network. \acp{DAS} make use of \acp{RRU} to extend the base stations' central antenna ports (cf. BS\,1 in \cref{fig:Network}). Coherent data transmission from all antennas is enabled by a high-bandwidth low-latency connection between the \acp{RRU} and the base station, thus making spatial multiplexing of several data-streams and/or users possible. Several publications have established the theoretical potential of \acp{DAS} for improving the network capacity, reducing the outage probability and improving the area spectral efficiency (e.g. \cite{Heath2011}), but investigations taking into account the constraints imposed by a practical system, e.g., channel coding, limited feedback, are scarce in literature (e.g.~\cite{Schwarz-DAS,Schwarz_DAS_Journal}). 

Simulations of advanced transceivers, especially for \ac{MUMIMO} transmission, as well as limited feedback algorithms require detailed knowledge of the users' channels, and are thus hardly amenable to system-level abstraction. Therefore link-level simulations appear as the appropriate choice, but are complicated by the fact that multiple base stations should be simulated to account for the changes in the out-of-cell interference environment caused by \acp{RRU}. In our simulations, we strike a compromise between computational complexity and accuracy, by employing the link-level simulator to explicitly simulate three cells and considering interference from more distant base stations with the out-of-cell interference model of \cite{Heath2011}. For that purpose, the Vienna \mbox{LTE-A} link-level simulator is extended with a distance-dependent pathloss model and a macroscopic fading model, determining the \ac{SNR} of a user based on its position (see \cite{Schwarz-DAS} for details).   

\cref{fig:ASE} shows simulation results obtained with this extended simulation environment. A network of base stations arranged in a regular hexagonal grid is considered. Each base station employs 120° sectorized transmit antennas and thus serves three cells. Additionally, each cell contains two \acp{RRU} at a distance of $2/3$ the cell radius (see BS~1 in \cref{fig:Network}). The throughput performance with and without \acp{RRU} is shown in the left and right parts of \cref{fig:ASE}, respectively. Without \acp{RRU} $N_t = 8$ transmit antennas are collocated at the base station, and with \acp{RRU} two antennas are placed at each \ac{RRU} leaving four collocated antennas for the base station. Each receiver is equipped with $N_r = 4$ antennas. 

In the simulations, different \ac{SUMIMO} and \ac{MUMIMO} transceivers are compared, assuming either perfect or quantized \ac{CSIT}. In \ac{SUMIMO}, users are served on separate time/frequency resources, thus avoiding in-cell interference between users. In a \ac{MUMIMO} system, users can additionally be multiplexed in the spatial domain. In this case, interference can be avoided by appropriate pre-processing at the transmitter, e.g., employing the simple linear precoding technique known as \ac{ZF} beamforming. The advantage of such techniques is that the potential spatial multiplexing gain is only limited by the number of transmit antennas, whereas in \ac{SUMIMO} the minimum of $N_t$ and $N_r$ is the decisive factor. Additionally, high receive antenna correlation often limits the spatial multiplexing capabilities of handheld devices strictly below $N_r$, a problem that is totally circumvented in \ac{MUMIMO} because different users typically experience uncorrelated channel conditions. But there is also a downside to \ac{MUMIMO}: Perfect interference-cancellation is only achieved with perfect \ac{CSIT}, otherwise residual interference impairs the transmission. Note that we restrict the \ac{MUMIMO} system to transmit at most one stream per user for simplicity. 

With perfect \ac{CSIT}, \cref{fig:ASE} shows that \ac{ZF} beamforming based \ac{MUMIMO} outperforms \ac{SUMIMO} with capacity achieving \ac{SVD} based transceivers, as soon as there are enough users per cell ($ \geq 8$) to exploit the spatial multiplexing capabilities of the base station.  Considering quantized \ac{CSIT}, \ac{ZF} beamforming performs similar to \ac{LTE}'s \ac{CLSM} in the centralized system, while in the \ac{DAS} a large throughput gain is achieved. This gain is enabled by investing the available feedback bits in those antennas of the distributed antenna array that experience a small macroscopic pathloss, thus exploiting the marco-diversity of the \ac{DAS}. On the other hand, \ac{MUMIMO} with unitary precoding based on \ac{PU2RC} achieves a lower throughput than \ac{SUMIMO} in both systems, because it cannot exploit the macro-diversity. Note that all considered algorithms have the same feedback overhead (an 8\,bit quantization codebook is used in all cases). For details on the considered transceiver architectures and feedback algorithms the interested reader is referred to~\cite{Schwarz_DAS_Journal}.  




\subsection{Macro-Femto Overlay Networks}
\label{sec:Martin}

One of the most efficient methods to enhance capacity in a macro cellular network is to minimize the distance between transmitter and receiver~\cite{Haenggi:2009}. This can be realized with smaller cell sizes and achieves the twofold benefits of increased spatial reuse and improved link quality. However, it comes at the cost of additional interference and required infrastructure. Femto cells are user-deployed low-power base stations, which offer an economical way to realize small cells in existing macro cellular networks. Since they primarily belong to the unplanned part of the network, it is one of the network providers' major concerns, how the link quality of macro cell attached users is influenced by a femto cell deployment. We investigate this question by enforcing two approaches:

\begin{enumerate}[leftmargin = 1.25em]
\item \textbf{Stochastic system model}: In order to carry out system-level simulations, it has been agreed in standardization meetings on models, such as the dual-stripe or the $5 \times 5$ grid model \cite{3gpp-RAN4-092042}. Although these models improve reproducibility, they do not meet scientific researchers' claim for analytical treatability. For this reason, we incorporate femto cell deployments in our \ac{LTE} system-level simulator based on \emph{stochastic geometry}. We utilize Poisson point- as well as Poisson cluster processes, which not only reflect the opportunistic placement of femto cells, but also provide analytically tractable expressions for performance metrics like outage probability and SINR~\cite{Dhillon:2011,Haenggi:2009}. The results obtained with these "simple" models indicate the same trends as the more elaborated environments mentioned above.

\begin{figure}
	\centering
	\includegraphics[scale = 0.8]{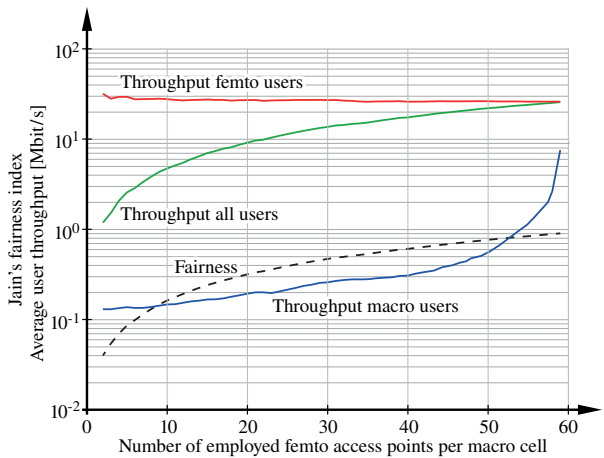}
	\caption{Comparison of the average user throughput achieved by macro cell users, femto cell users and both types combined versus the number of femto access points per macro cell. Also shown is the fairness of the resource allocation in terms of Jain's fairness index.}
	\label{Fig:ThroughputAndFairness}
\end{figure}

\item \textbf{Fairness metric}: Based on the stochastic models, we investigate \emph{how many femto cells can be beneficially deployed in an existing macro cell}, which arises the prior question: \emph{Beneficial in which sense?}

In our setup, clusters of users are spread homogeneously over the macro cell area (according to a Poisson cluster process). Then, one by one, femto cell access points are added to the network and placed at the center of these clusters. Thus, an increasing amount of users is in coverage of a femto cell while the total number of users remains constant. \cref{Fig:ThroughputAndFairness} depicts simulation results for the average user throughput (middle solid line) plotted versus the femto cell density, i.e., the number of employed femto cell access points per macro cell. The curve indicates performance improvements for an increasing number of femto cells.  However, it conceals the imbalance of femto cell- and macro cell user throughput, as shown by the upper- and lower solid line in \cref{Fig:ThroughputAndFairness}, respectively. Therefore, we emphasize the importance of a fairness metric to quantify the distribution of the throughput values among the users. In our work, we utilize Jain's fairness index, as shown by the dashed line in the figure.

\end{enumerate}

Succinctly, we stress the significance of incorporating various metrics into the performance assessment of heterogeneous cellular networks, and encourage to apply stochastic geometry for the system models of such networks.


\section{Conclusions}
\label{sec:Conclusions}

In this article, we present our approach to bring researchers and standardization/industry experts in LTE closer together, by means of the \textit{Vienna LTE simulators}, a standard-compliant open-source \mat-based simulation platform for 3GPP LTE and LTE-A. The simulators facilitate a standard-compliant validation of novel research results and simplify the evaluation of such results in terms of their significance for practical systems. Furthermore, a unifying platform greatly improves the reproducibility and comparability of different algorithms, by removing uncertainties in the multitude of simulation parameters.    

We demonstrate the capabilities of this approach by giving an overview of the different research directions pursued in our research group with the aid of the simulators. Topics such diverse as pilot-power allocation and frequency synchronization, multi-user scheduling and beamforming, and heterogenous network architectures can be effectively treated and investigated with the simulator platform. 

We thus encourage researches and engineers, whose field of work is related to LTE/LTE-A, to take a closer look at the \textit{Vienna LTE simulators} and find out whether they can benefit from using the platform.

\section*{Acknowledgments}
This work has been funded by the Christian Doppler Laboratory for Wireless Technologies for Sustainable Mobility, KATHREIN-Werke KG, and A1 Telekom Austria AG. The financial support by the Federal Ministry of Economy, Family and Youth and the National Foundation for Research, Technology and Development is gratefully acknowledged. The authors would like to thank Christoph~F. Mecklenbr\"auker, Robert~W. Heath~Jr. and Paulo~S.~R.~Diniz for valuable comments and encouraging discussions. We appreciate the helpful feedback and comments from the online-community that has evolved around the \textit{Vienna LTE simulators}.

\bibliographystyle{IEEEtran_no_url}

\begin{footnotesize}
\bibliography{standards_short,commag}

\begin{thebibliography}{10}
\providecommand{\url}[1]{#1}
\csname url@rmstyle\endcsname
\providecommand{\newblock}{\relax}
\providecommand{\bibinfo}[2]{#2}
\providecommand\BIBentrySTDinterwordspacing{\spaceskip=0pt\relax}
\providecommand\BIBentryALTinterwordstretchfactor{4}
\providecommand\BIBentryALTinterwordspacing{\spaceskip=\fontdimen2\font plus
\BIBentryALTinterwordstretchfactor\fontdimen3\font minus
  \fontdimen4\font\relax}
\providecommand\BIBforeignlanguage[2]{{%
\expandafter\ifx\csname l@#1\endcsname\relax
\typeout{** WARNING: IEEEtran.bst: No hyphenation pattern has been}%
\typeout{** loaded for the language `#1'. Using the pattern for}%
\typeout{** the default language instead.}%
\else
\language=\csname l@#1\endcsname
\fi
#2}}

\bibitem{ViennaLTESimulators}
C.~Mehlf\"uhrer, J.~C. Ikuno, M.~Simko, S.~Schwarz, and M.~Rupp, ``The {V}ienna
  {LTE} simulators --- {E}nabling {R}eproducibility in {W}ireless
  {C}ommunications {R}esearch,'' \emph{{EURASIP} {J}ournal on {A}dvances in
  {S}ignal {P}rocessing ({JASP}) special issue on Reproducible Research}, 2011.

\bibitem{Caban2012}
S.~Caban, C.~Mehlf{\"u}hrer, M.~Rupp, and M.~Wrulich, Eds., \emph{Evaluation of
  {HSDPA} and {LTE:} From Testbed Measurements to System Level
  Performance}.\hskip 1em plus 0.5em minus 0.4em\relax {UK}: John Wiley {\&}
  Sons, 2012.

\bibitem{VTC2011_fall_Michal}
M.~{\v S}imko, S.~Pendl, S.~Schwarz, Q.~Wang, J.~C. Ikuno, and M.~Rupp,
  ``{O}ptimal {P}ilot {S}ymbol {P}ower {A}llocation in {LTE},'' in \emph{Proc.
  74th IEEE Vehicular Technology Conference (VTC2011-Fall)}, San Francisco,
  USA, Sept. 2011.

\bibitem{eurasip_michal_2012}
M.~{\v S}imko, Q.~Wang, and M.~Rupp, ``{O}ptimal {P}ilot {S}ymbol {P}ower
  {A}llocation under {T}ime-variant {C}hannels,'' \emph{{EURASIP} {J}ournal on
  {W}ireless {C}ommunications and {N}etworking}, 2012.

\bibitem{VTC2012_fall_Michal}
M.~{\v S}imko, P.~S.~R. Diniz, Q.~Wang, and M.~Rupp, ``{P}ower {E}fficient
  {P}ilot {S}ymbol {P}ower {A}llocation under {T}ime-variant {C}hannels,'' in
  \emph{Proc. 76th IEEE Vehicular Technology Conference (VTC2011-Fall)},
  Quebec, Canada, Sept. 2012.

\bibitem{PIMRC2010_Qi}
Q.~Wang, C.~Mehlf\"uhrer, and M.~Rupp, ``Carrier frequency synchronization in
  the downlink of {3GPP} {LTE},'' in \emph{Proceeding of the 21st Annual {IEEE}
  International Symposium on Personal, Indoor and Mobile Radio Communications
  ({PIMRC}'10)}, Istanbul, Turkey, Sept. 2010.

\bibitem{Asilomar2011_Qi}
Q.~Wang and M.~Rupp, ``Analytical link performance evaluation of {LTE} downlink
  with carrier frequency offset,'' in \emph{Conference Record of the
  Fourtyfifth Asilomar Conference on Signals, Systems and Computers, 2011
  ({Asilomar-2011})}, Pacific Grove, USA, Nov. 2011.

\bibitem{heath2001multiuser}
R.~Heath~Jr, M.~Airy, and A.~Paulraj, ``Multiuser diversity for {MIMO} wireless
  systems with linear receivers,'' in \emph{Conference Record of the
  Thirty-Fifth Asilomar Conference on Signals, Systems and Computers,
  2001.}\hskip 1em plus 0.5em minus 0.4em\relax IEEE, 2001.

\bibitem{ikuno2010system}
J.~Ikuno, M.~Wrulich, and M.~Rupp, ``System level simulation of {LTE}
  networks,'' in \emph{2010 IEEE 71st Vehicular Technology Conference (VTC
  2010-Spring)}.\hskip 1em plus 0.5em minus 0.4em\relax IEEE, 2010.

\bibitem{Heath2011}
R.~{Heath Jr.}, T.~Wu, Y.~H. Kwon, and A.~Soong, ``Multiuser {MIMO} in
  distributed antenna systems with out-of-cell interference,'' \emph{IEEE
  Transactions on Signal Processing}, vol.~59, no.~10, Oct. 2011.

\bibitem{Schwarz-DAS}
S.~Schwarz, R.~{Heath Jr.}, and M.~Rupp, ``Multiuser {MIMO} in distributed
  antenna systems with limited feedback,'' Anaheim, Ca., Dec. 2012, submitted
  to IEEE 4th Int. Workshop on Heterogeneous and Small Cell Networks, GLOBECOM
  2012.

\bibitem{Schwarz_DAS_Journal}
S.~Schwarz, R.~{Heath, Jr.}, and M.~Rupp, ``Single-user {MIMO} versus
  multi-user {MIMO} in distributed antenna systems with limited feedback,''
  \emph{Submitted for review to {EURASIP} {J}ournal on {A}dvances in {S}ignal
  {P}rocessing ({JASP})}, 2012.

\bibitem{Haenggi:2009}
M.~Haenggi and R.~K. Ganti, ``Interference in large wireless networks,''
  \emph{Found. Trends Netw.}, vol.~3, no.~2, pp. 127--248, Feb. 2009.

\bibitem{3gpp-RAN4-092042}
{3GPP RAN4}, ``{Simulation Assumptions and Parameters for FDD HeNB RF
  Requirements},'' March 2009.

\bibitem{Dhillon:2011}
H.~S. Dhillon, R.~K. Ganti, F.~Baccelli, and J.~G. Andrews, ``Modeling and
  analysis of k-tier downlink heterogeneous cellular networks,'' \emph{CoRR},
  vol. abs/1103.2177, 2011.

\end{thebibliography}
\end{footnotesize}

\end{document}